\documentclass[prd,preprint,scriptaddress]{revtex4}
\input{epsf}
\newcommand{\ba}{\begin{array}}
\newcommand{\ea}{\end{array}}
\newcommand{\bd}{\begin{displaymath}}
\newcommand{\ed}{\end{displaymath}}
\newcommand{\be}{\begin{equation}}
\newcommand{\ee}{\end{equation}}
\newcommand{\bea}{\begin{eqnarray}}
\newcommand{\eea}{\end{eqnarray}}

\def\L{\Lambda}

\def\Rm{\rho_m}
\def\Rx{\rho_x}
\def\r+{r_s^+}
\def\r-{r_s^-}
\begin{document}
\title{CMB constraints on interacting cosmological models}
\author{Diego Pav\'{o}n\footnote{electronic addres: diego.pavon@uab.es}}
\affiliation{Departamento de Fisica, Universidad Aut\'{o}onoma de Barcelona,
08193 Bellaterra (Barcelona), Spain.}
\author{Somasri Sen\footnote{electronic addres: somasri@cosmo.fis.fc.ul.pt}}
\affiliation{CAAUL, Departamento de Fisica da FCUL, Campo Grande, 1749-016
Lisboa, Portugal.}
\author{Winfried Zimdahl\footnote{electronic addres: zimdahl@thp.uni-koeln.de}}
\address{Institut f\"ur Theoretische Physik, Universit\"{a}t zu
K\"{o}ln, 50937 K\"{o}ln, Germany.}

\begin{abstract}
Non--canonical scaling of the form $\rho_x\propto \rho_m
a^\xi$, where $\rho_x$ and $\rho_m$ are the energy densities of
dark energy and dark matter, respectively, provides a natural way
to address   the coincidence problem -why the two densities are 
comparable today. This non--canonical scaling is achieved by a
suitable interaction between the two dark components. We study the
observational constraints imposed on such models from the cosmic
microwave background (CMB) anisotropy spectrum. Using the recent
WMAP results for the location of the CMB peaks, we determine the
admissible parameter space and find that interacting models are
well consistent with current observational bounds.
\end{abstract}

\maketitle

\section{Introduction}
The picture that emerges from the present cosmological
observations portrays a spatially flat, low matter density
universe, currently undergoing a stage of accelerated expansion
\cite{cmbsuper}. Within Einstein's relativity, the simplest
explanation for this acceleration requires two dark components:
one in the form of non--luminous dust (``dark matter'')  with
negligible pressure, contributing roughly  one-third of the
total energy density  of the Universe and clustering
gravitationally at small scales, and the  other one a smoothly
distributed component having a large negative pressure (``dark
energy'') and contributing about two--thirds of the total energy
density of the Universe. Although the immediate candidate for this
dark energy is the vacuum energy (i.e., a cosmological constant
$\Lambda$), alternative scenarios, with acceleration driven by a
dynamical scalar field, called ``quintessence'', were proposed
from different perspectives \cite{diff}.

One problem that afflicts most of the models proposed so far is
the so--called ``{\em coincidence problem}'' or ``{\em why now?}''
problem \cite{stein}. The essence of this problem is as follows:
as these two dark components redshift with expansion at different
rates one may ask ``{\em why the ratio of these two components is
of the same order precisely today?}, i.e., {\em why
$\frac{\Rm}{\Rx}|_0={\cal O} (1)$ ?}'' There have been different
approaches for solving it \cite{anek,diego1}.  The one  to be
discussed here considers some non--canonical scaling of the ratio
of  dark matter and dark energy  with the Robertson--Walker
scale factor $a(t)$ \cite{Dalal, scaling}
\\
\be \frac{\Rm}{\Rx} \propto a^{-\xi} \  , \ee where $\Rm$ and
$\Rx$ are the energy densities of the dark matter and the dark
energy, respectively, and the scaling parameter $\xi$ is a new
quantity which assesses the severity of the problem. Thus, $\xi=3$
corresponds to the $\L$CDM model and  $\xi=0$ to the self similar
solution with no coincidence problem. Hence, any solution with a
scaling parameter $0<\xi<3$ makes the coincidence problem less
severe. 
   The standard noninteracting cosmology is characterized by 
the relation $\xi=-3w_x$, where $w_{x} \equiv p_{x}/\Rx$ is the 
equation of state parameter 
for the dark energy which we here assume to be 
constant for simplicity.   
Solutions deviating from    this    relation represent 
a testable, nonstandard cosmology. 
Such deviation from standard dynamics can be
obtained if the two dark components are not separately conserved
but coupled to each other. This proposal has been explored in
\cite{scaling, plb} and looks promising as a suitable mutual
interaction can make both components redshift coherently.

Now,  it seems advantageous to use the available
observational information to constrain the nature of dark energy
with minimal theoretical input rather than to perform a detailed
fit to a particular model with a specific potential.  This
approach has been followed in \cite{Dalal}, where high redshift
supernovae survey, Alcock--Paczynski test to quasar pairs and
evolution of cluster abundances have been used to constrain the
scaling parameter $\xi$ and the dark energy equation of state parameter
$w_x$ for the case of separately conserved components. As shown in
\cite{scaling} the    high redshift SNIa data given in \cite{perl}
cannot discriminate between  interacting models and the ``concordance"
$\Lambda$CDM model. Therefore, the
target of this work is to constrain the parameters $\xi$ and $w_x$ 
for this kind of non--standard cosmology  with the Cosmic Microwave 
Background (CMB) anisotropy spectrum. To this end, we use the Wilkinson
Microwave Anisotropy Probe (WMAP) \cite{wmap} and BOOMERanG
\cite{boom} data for the location of the peaks in the angular
spectrum. As it turns out, the scaling interacting cosmology is 
well consistent with the observational bounds.   
The parameter space corresponding to interacting models, seems  
even to be favoured compared with the parameter space of the standard 
noninteracting models.   

Section 2 succinctly recalls and slightly generalizes, by 
introducing a radiation component,  
the scaling model of Ref.\cite{scaling};  
this component is    crucial, of course,    if one wishes to 
   test the interacting cosmology       
with the CMB data. In section 3 the positions of the CMB 
peaks as witnessed by WMAP are used to constrain 
the model. Section 4 summarizes our findings. Finally, the Appendix
collects the set of formulae employed in our analysis.

\section{Scaling Solutions}
We consider a  Friedmann--Lema\^{\i}tre--Robertson--Walker (FLRW)
universe such that its total energy density consists of radiation 
(whose energy density lies at present by four orders of magnitude 
below that of matter), matter with negligible pressure (that 
encompasses both baryonic and non--baryonic components) 
and dark energy, 
\\
\be
\rho = \rho_r+\Rm+\Rx \ .
\ee
The corresponding pressures are
\\
\be
p_r = \textstyle{1\over{3}}\rho_r, \quad p_m<<\Rm, \quad p_x= w_x\Rx \ .
\ee

The $m$ and $x$ components are supposed to share some
coupling so that their energy densities obey the balances
\\
\bea
\dot\rho_r+4H\rho_r=0 \ , \\
\dot\Rm+3H\Rm=Q \, , \nonumber\\
\dot\Rx+3H(1+w_x)\Rx=-Q \ , \eea
\\
where $H \equiv \dot a/a$ is the Hubble factor and the
(non-negative) quantity $Q$ measures the strength of the
interaction.  For simplicity this setup is chosen such that the
entire matter component takes part in the coupling. Alternatively,
one may treat the baryonic component as separately conserved. The
constraints obtained below do not depend on whether or not the
baryons are included in the interaction. 

From Eq. (4) it follows that the radiation redshifts as
\\
\be \rho_r=\rho_{r0}\left(\frac{a_0}{a}\right)^{4}\ ,
\ee
where $a_{0}$ is the  current  value of the scale factor. 

We are interested in solutions with the following scaling behaviour for
the two dark components,
\be
\frac{\Rm}{\Rx}=r_0\left(\frac{a_0}{a}\right)^\xi\ ,
\ee
\noindent
where $r_0$ denotes the ratio of both components at present time and
$\xi$ is a constant parameter in the range $0 \leq \xi \leq 3$. 
In \cite{scaling}, it was shown that such scaling solutions follow
from an interaction characterized by   
\\
\be Q=- 3H \, \frac{(\xi/3)+w_x}{1+r_0(1+z)^\xi}\, \Rm \
, \label{8}
\ee
\noindent
where $1+z = a_{0}/a(t)$. For the standard cosmology without
interaction - i.e., when $Q=0$ -, we  have $\xi=-3 w_x$.
\noindent The $\Lambda$CDM model $(w_{x}=-1)$ is the special case
with $\xi=3$. Any deviation from $Q=0$, i.e, $(\xi/3)+ w_x=0$,
implies an alternative, testable, non--standard cosmology.
With the interaction (\ref{8}) it is straightforward to find
\\
\be \Rm+\Rx=\rho_{m0} \, \frac{1+r_0}{r_0}(1+z)^{3(1+w_x)}
\left[\frac{1+r_0(1+z)^\xi}{1+r_0}\right]^{-\frac{3 w_x}{\xi}}\ .
\ee Again, for $w_x=-1$ and $\xi=3$ this reduces to
\noindent
\be
\Rm+\Rx=\frac{\rho_{m0}}{r_0}[1+r_0(1+z)^3]\ ,
\ee
which is indeed the prediction of the $\Lambda$CDM model.
\\
\noindent 
With the evolution of the components as given in Eqs.(6)
and (9), the Friedmann equation can be written as
\\
\bea
H^2&=&\frac{8\pi G}{3}\rho \nonumber\\
&=&H_0^{2} \, \left[\Omega_{m0}\frac{1+r_0}{r_0}(1+z)^{3(1+w_x)}
\left[\frac{1+r_0(1+z)^\xi}{1+r_0}\right]^{-\frac{3
w_x}{\xi}}\right.\nonumber\\
&& +\left.\Omega_{ro}(1+z)^4\right] \ ,\eea
\\
where we have introduced the dimensionless density parameters \be
\Omega_{m0}=\frac{\rho_{m0}}{\rho_c}\ ,\quad \Omega_{x0}
=\frac{\rho_{x0}}{\rho_c}\ , \quad
\Omega_{r0}=\frac{\rho_{m0}}{\rho_c}\ , \quad \rho_c \equiv
\frac{3H_0^2}{8\pi G} \ . \ee \noindent We can recast the last
equation into
\\
\be
H^2=\Omega_{m0}H_0^2(1+z)^{4} \; X(z) \  ,
\ee
with
\be
X(z)=\frac{1+r_0}{r_0} \, (1+z)^{3(1+ w_x)-4}
\left[\frac{1+r_0(1+z)^\xi}{1+r_0}\right]^{-\frac{3 w_x}{\xi}}
+\frac{\Omega_{r0}}{\Omega_{m0}}\ .
\ee
The above form of $H^2$ is useful in testing various
models against CMB observation \cite{card}. \\

\section{Constraints from CMB}
The CMB acoustic peaks and troughs arise from oscillations of the
primeval plasma just before the Universe becomes translucent. 
These oscillations are the result of a balance between the
gravitational interaction and the photon pressure in the tightly
bound photon-baryon fluid. The locations of the peaks
corresponding to different angular momenta depend on the acoustic
scale $l_A$, which in turn is related to the angular diameter
distance $D$ to the last scattering, and on the sound horizon
$s_{ls}$ at last scattering through $l_A = \pi D/s_{ls}$
\cite{hu}. To a good approximation this ratio for $l_A$ is
\cite{doran1} 
\\
\be
l_A=\pi\frac{\tilde\tau_0-\tilde\tau_{ls}}{\bar{c_s}\tilde\tau_{ls}}
\ , 
\ee
where $\tilde\tau(=\int a^{-1}dt)$ is the conformal time
and the subscripts $0$ and $ls$ represent the time at present and
at the last scattering era, respectively. $\bar{c_s}$ is the
average sound speed before last scattering defined by \be
\bar{c_s}\equiv \tilde\tau_{ls}^{-1}\int_0^{\tilde\tau_{ls}} c_s
d\tilde\tau, \ee with \be
c_{s}^{-2}=3+\textstyle{9\over{4}}\frac{\rho_b}{\rho_r} \, , \ee
where $\rho_{b}$ stands for the energy density of baryons.

In an ideal photon-baryon fluid model, the  analytic relation
between the position $l_m$ of the m-th peak and the acoustic scale
$l_A$ is $l_m=m~l_A$.  But this simplicity gets disturbed by the
different driving and dissipative effects which  induce a shift
with respect to the ideal position \cite{hu}. This shift has
been  accounted for by parametrizing the location of the peaks
and troughs by
\\
\be
l _m\equiv l_A(m-\phi_m)\equiv l_A(m-{\bar\phi}-\delta\phi_m)
\ ,
\ee
where ${\bar\phi}$ is  an overall peak shift, identified with
$\phi_{1}$,  and $\delta\phi_m\equiv\phi_m-{\bar\phi}$ is the
relative shift of the m-th peak. This parametrization can be used
to extract information about the matter content of the Universe
before last scattering. Although it is certainly very difficult to
derive an analytical relation between cosmological parameters and
phase shifts, Doran and Lilley \cite{doran2} have given certain
fitting formulae which makes life much simpler. These formulae do
not have a prior and crucially depend on cosmological parameters
like spectral index $(n_s)$, baryon density $(\omega_{b}=\Omega_b
h^2)$, (normalized) Hubble parameter $(h)$, ratio of
radiation to matter at last scattering $(r_{ls})$ and also  on
$\Omega^d_{ls}$, representing the dark energy density at the
time of recombination. We use these formulae (collected in the
Appendix) to specify the positions of the peaks in the scaling
model and constrain its parameter space with the WMAP data.

It should be mentioned here that although these formulae were
obtained for quintessence models with an exponential potential,
they are expected to be fairly independent of the form of the
potential and the nature of the late time acceleration mechanism,
as shifts are practically independent of post recombination
physics. It should also be stressed that the errors associated
with the analytical estimators for the peak positions we are
using, determined in comparison with the CMBFAST for standard
models, is less than 1\% \cite{doran2}.

Let us now turn to discussing the latest available CMB data on the
positions of the peaks. The bounds on the locations of the first
two acoustic peaks and the first trough from the WMAP measurement
\cite{wmap} of the CMB temperature angular power spectrum are \bea
l_{p_1} &=& 220.1\pm 0.8,\nonumber\\
l_{p_2} &=& 546\pm 10,\\
l_{d_1} &=& 411.7\pm 3.5;\nonumber \eea notice that all
uncertainties are within $1\sigma$ and include calibration and
beam errors. The location for the third peak is given by BOOMERanG
measurements \cite{boom} \be l_{p_3}=825^{+10}_{-13}\ .
\ee

Now, with the help of the above formulae  we calculate the position of
the peaks in the CMB spectrum in scaling models and constrain the model
parameters to the values consistent with the observational bounds.
The acoustic scale is determined in terms of the conformal time $\tilde{\tau}$
in Eq. (15). From Eq.(13) we get
\\
\be \tilde\tau_{ls}=\int_0^{\tilde\tau_{ls}} d\tilde\tau=
\frac{1}{\Omega_{m0}^{1/2}H_0}\int_0^{a_{ls}}\frac{da}{X(a)} \  ,
\ee and \be \tilde\tau_{0}=\int_0^{\tilde\tau_{0}} d\tilde\tau=
\frac{1}{\Omega_{m0}^{1/2}H_0}\int_0^1 \frac{da}{X(a)} \  , \ee
where $X(a)$ is given by Eq.(14) and we have chosen $a_0 =1$.

Inserting the above expression in equation (15), we obtain an
analytical expression for $l_A$, 
\\
\be
l_A=\frac{\pi}{\bar{c_s}}\left[\frac{\int_0^1
\frac{da}{X(a)}}{\int_0^{a_{ls}} \frac{da}{X(a)}}-1\right] \ .
\ee

From the computation of the acoustic scale by Eq. (23), the
equations for the peak shifts Eq. (18), and the fitting formulae
in the Appendix, we look for the combination of the model
parameters that are consistent with the observational bounds. We
have plotted the contours consistent with the bounds of the first
three acoustic peaks and the first trough corresponding to the
WMAP and BOOMERanG data given by Eqs. (19) and (20) in the
$\xi-w_{x}$ parameter space for different values of $n_s$ and
$\Omega_{m0}$. The  investigated cosmological parameter space 
is given by $(n_{s}, h, \omega_b, \Omega_m)$. Throughout this
paper, we have neglected the contribution from the spatial
curvature and massive neutrinos. We have also neglected the
contributions from  gravitational waves in the initial
fluctuations. Because of the rather tight WMAP constraint on
$\omega_b,(\omega_b=0.0224\pm 0.0009$ \cite{wmap}), we have
assumed $\omega_b=0.0224$ in our calculations. To have a clear
idea of the dependence on the parameters we have drawn  two
different types of  plots.  The first three figures  show the $\xi-w_{x}$ 
parameter space for a particular value of $h$($=0.71$) 
and different values of $n_s$ and $\Omega_{m0}$. The next
two figures depict the $\Omega_{m0}-h$ parameter space for
different values of $n_s$, $\xi$ and $w_{x}$.

As already mentioned, we have assumed in the general outline of
Sec.2 that the entire matter component, including the baryons,
takes part in the interaction with the dark energy. We have also
investigated the case where only the non--baryonic matter
part is coupled to the dark energy, while the baryonic
energy density is locally conserved. The resulting plots do not
depend on these different ways to implement the interaction. 

In figure 1, we have plotted the contours corresponding to the
bounds of the first three  peaks and the first dip (Eqs. (19)
and (20)) in the $\xi-w_{x}$ parameter space with $h=0.71$ and
$\Omega_{m0}=0.2$ for different values of $n_s$. Figures 2 and 3
represent similar contours with same $h$ and $n_s$ but different
$\Omega_{m0}$ (0.3 and 0.4, respectively). To facilitate the
analysis, all these figures show the non--interaction line,
$\xi+3w_{x}=0$. It is apparent that the interaction is severely
restricted  by the CMB data.  And simultaneously it is very 
interesting to notice that in Figs 2 and 3 the line $\xi + 3w_x=0$ 
is clearly outside the allowed CMB regions which    implies that
with the assumed  priors ($h=0.71$ and the mentioned value of $n_s$ 
and $\Omega_{m0}$) there is no parameter space at all consistent 
with a noninteracting cosmology, including the $\Lambda$CDM model.

Since the above important conclusion depends crucially on the priors 
chosen, we investigate the same models in a different parameter space.
Figure 4 depicts the contours for the same peaks and dips in the
$\Omega_{m0}-h$ parameter space with $n_s=0.97$ and with three 
different values of $\xi$ and $w_{x}$. We have chosen the 
values of $\xi$ as $1.0,2.0,3.0$.
The higher the value of $\xi$, the more acute the coincidence problem.
For $w_{x}$ we have chosen $-0.5,-0.75, -1.0$. Of course, for $\xi=3.0$ and
$w_{x}=-1.0$ the model collapses to the $\L$CDM model.
Figure 5 represents the same contour as figure 4 with
$n_s=1.0$ and the same values for $\xi$ and $w_{x}$.

\section{Discussion}
We have investigated the constraints imposed by the observed
positions of the peaks of the CMB anisotropy spectrum -as
witnessed by WMAP and the BOOMERanG experiments \cite{wmap},
\cite{boom}-, on the non--standard scaling interacting cosmology
$\xi+3w_{x}\neq 0$ model \cite{scaling}.  The most important
consequence of our analysis is that the non--standard scaling
interacting cosmology shows good consistency with the observational
bounds for a variety of parameter combinations.  
   The parameter space favoured by the CMB is larger for interacting 
cosmological models than for  noninteracting ones.   

In Fig.1, for $\Omega_{m0}=0.2$, the CMB data suggests a constrained 
parameter space for all the four values of $n_s$, whereas in Fig.2, for 
$\Omega_{m0}=0.3$, we get    an allowed    region only for $n_s\leq 0.97$. 
In Fig 3, for $\Omega_{m0}=0.4$, we do not have an admissible   parameter 
space at all. In Figs. 1 and 2, $w_x$ varies in the limits $-0.38>w_x>-1.08$ over 
the entire range of $\xi  (0\leq\xi\leq 3)$. Thus, from the whole set of 
figures it is rather obvious that the scaling interacting model favours 
lower value of $n_{s}$ and a moderate range of values for $\Omega_{m0}$. 
On the other hand, the noninteracting model satisfies the parameter space 
bounded by CMB data only in Fig 1., i.e, for $\Omega_{m0}=0.2$. In  
figures 2 and 3 the noninteraction line stays well beyond the reach 
of the parameter space allowed by the CMB data. This includes the 
concordance $\Lambda$CDM model as well. Thus, the scaling model is better 
consistent with the CMB data and it is compatible with a larger 
parameter space than  the noninteracting standard model. 

It is worthy of note
that for $\xi<1$, the region bounded by the contours of Figs. 1, 2
and 3 reduces practically to a line. For the stationary solution
with no coincidence problem, $\xi\simeq 0$, we practically get a
particular value for $w_{x}$ satisfied by the bounds of WMAP and
BOOMERanG data. This value of $w_{x}$ varies only with
$\Omega_{m0}$  -between $w_{x} \simeq -0.38$, when
$\Omega_{m0} = 0.2$, and $w_{x} \simeq -0.52$, when $\Omega_{m0} =
0.4$. (It should be borne in mind that claims according to which
observations imply that $w_{x} < -0.87$ \cite{melchiorri} are
based in non--interacting cosmologies). 

As mentioned above in figures 1, 2 and 3, 
the constrained region crucially depends on $h$ and $n_s$.  To
make this clear, we present 
in Figs.4 and 5 contours in the $\Omega_{m0}-h$ parameter space 
for the same $\xi~{\rm and}~w_{x}$ but for 
different values of $n_s (0.97,1.0)$. From these plots it
becomes  also apparent that for a lower value of $\xi$, i.e., 
for a less severe coincidence problem, we have  an admissible $\Omega_{m0}-h$ 
parameter space for higher values (lower amounts) of the equation of 
state parameter $w_{x}$.  For example in the first column of panels in Figs. 4 and 5 
(for $\xi=1.0$) we have an allowed parameter space for $w_x=-0.50$ 
only, for $w_x=-0.75$ there is no admissible   space and for $w_x=-1.0$ 
the contours go beyond the range of the parameter space chosen. 
Similarly, as the coincidence problem  becomes more acute, i.e., 
as $\xi$ grows, lower values (higher amounts)   of $w_{_x}$ become more favoured.  
This implies that thanks to the interaction 
the dark energy pressure becomes less negative.
The bottom-right panel in both figures corresponds to the $\Lambda$CDM model 
$(\xi=3.0~{\rm and}~ w_x=-1.0)$. For the panel in Fig. 4 representing 
the $\Lambda$CDM model the CMB bounds are satisfied for $0.19<\Omega_{m0}<0.3$ 
and $h\geq 0.68$, while for the corresponding panel representing the $\Lambda$CDM model 
in Fig. 5, the bounds are satisfied only for $h>0.72$ and $\Omega_{m0}<0.22$. 
This depicts that $\Lambda$CDM favours low $n_s$ values. This result 
seems quite consistent with that of Ref. \cite{card}.

After the recent high redshift SNIa data \cite{tonry,barris}
lending additional support to an accelerating universe picture,
and with further strings of observations (like those from the SNAP
satellite) yet to come, one major worry of the community is placed
on the nature of dark energy \cite{paris}. In this respect, a
scaling solution of the type $\Rm/\Rx=r_0(a_0/a)^\xi$  appears to be
a quite promising tool for analyzing the relationship between the
two forms of energy dominating our present Universe. And with the
available constraints from WMAP and BOOMERanG experiments, we 
can certainly conclude that interacting cosmological models 
$(\xi+3 w_{x}\neq 0)$ may well compete  with the ``concordance"
$\Lambda$CDM model, they seem even to be favoured when 
compared to the latter one. 

While the present cosmological data are insufficient to discriminate
between these models, it is to be hoped that future observations of
high redshift SNIa as well as other complementary data (from galaxy
clusters evolution and lensing effects) will decisively help to break
the degeneracy.

\acknowledgments{This work was partially supported by by the Funda\c c\~ao
para a Ci\^encia e a Tecnologia (Portugal), through CAAUL, the
Spanish Ministry of Sceince and Technology under grant
BFM2003-06033, NATO and Deutsche Forschungsgemeinschaft.}

\appendix
\section*{Appendix}

For completeness, we put together here the formulas used in our search
for parameter space. These fitting formulae are quoted from the cited
literature \cite{doran2}.

We assume the standard recombination history and define the
redshift $z_{ls}$ of decoupling as the redshift at which the
optical depth of Thompson scattering is unity. A useful fitting
formula for $z_{ls}$ is given by \cite{param}: \be
z_{ls}=1048[1+0.00124 \omega_b^{-0.738}][1+g_1 \omega_m^{g_2}],
\label{zdec} \ee where
$$
g_1=0.0783 \omega_b^{-0.238}[1+39.5 \omega_b^{0.763}]^{-1},\;\;\;
g_2=0.56[1+21.1 \omega_b^{1.81}]^{-1},
$$
$\omega_b\equiv\Omega_bh^2$ and $\omega_m\equiv\Omega_mh^2$.\\

The ratio of radiation to matter at last scattering is
\begin{equation}\label{r_star}
r_{ls} = \rho_r(z_{ls}) / \rho_m(z_{ls}) =
0.0416 \omega_m^{-1} \left(z_{ls} / 10^3\right).
\end{equation}

The overall phase shift $\bar{\varphi}$ (which is  the phase
shift of the first peak) is parametrized by the formula
\begin{equation} \label{phi_1}
\bar{\varphi} =(1.466 - 0.466n_s) \left[ a_1 r_{{ls}}^{a_2} + 0.291
  \bar\Omega_{ls}^{d} \right],
\end{equation}
where $a_1$ and $a_2$ are given by
\begin{eqnarray}
a_1 & = & 0.286 + 0.626~ \omega_b   \, , \\
a_2 & = & 0.1786 - 6.308~ \omega_b + 174.9~ \omega_b^2 - 1168~ \omega_b^3 . 
\end{eqnarray}

The relative shift of the second peak $(\delta\phi_2)$ is given by
\begin{equation}
\delta \varphi_2 = c_0 - c_1 r_{ls} - c_2 r_{ls} ^{-c_3} + 0.05\,
(n_s-1),
\end{equation}
with
\begin{eqnarray}
c_0 &=& -0.1 + \left( 0.213 - 0.123 \bar\Omega_{ls}^d \right)\\
&&\times \exp\left\{ - \left( 52 - 63.6\bar\Omega_{ls}^d\right)
\omega_b \right\}\\
c_1 &=& 0.063 \,\exp \{-3500~ \omega_b^2 \} +
0.015\\
c_2 &=& 6\times 10^{-6} + 0.137 \left (\omega_b - 0.07 \right) ^2\\
c_3 &=& 0.8 + 2.3 \bar\Omega_{ls}^d +
\left( 70 - 126\bar\Omega_{ls}^d\right) \omega_b.
\end{eqnarray}

For the third peak we have,
\begin{equation}
\delta \varphi_3 = 10 - d_1r_{ls}^{d_2} + 0.08\, (n_s-1),
\end{equation}
with
\begin{eqnarray}
d_1 &=& 9.97 + \left(3.3 -3 \bar\Omega_{ls}^d\right) \omega_b \\
\nonumber d_2 &=& 0.0016 - 0.0067\bar\Omega_{ls}^d +
\left(0.196 - 0.22\bar\Omega_{ls}^d\right)\omega_b\\
&& + \frac{(2.25 + 2.77 \bar\Omega_{ls}^d) \times 10^{-5}}{\omega_b}.
\end{eqnarray}

The relative shift of the first trough is given by

\be
\delta\varphi_{3/2}=b_0+b_1r_{ls}^{1/3}\exp(b_2r_{ls})+0.158(n_s-1)~
\label{eq:delphi}
\ee
with

\bea
b_0 &=& -0.086-0.079 {\bar\Omega}_{ls}^{d}
       -\left( 2.22-18.1{\bar\Omega}^{d}_{ls} \right) \omega_b\nonumber\\
    &&-\left(140+403{\bar\Omega}^{d}_{ls} \right)\omega_b^2~,\nonumber\\
b_1 &=& 0.39-0.98{\bar\Omega}^{d}_{ls}
       -\left(18.1-29.2{\bar\Omega}_{d}^{ls}\right)\omega_b\nonumber\\
    && +440\omega_b^2,\\
b_2 &=& -0.57-3.8\exp({-2365\omega_b^2})~. 
\label{eq:bs}
\eea

The overall shifts for the second and the third peaks and for the first
trough are $\bar\phi+\delta\phi_2$, $\bar\phi+\delta\phi_3$ and
$\bar\phi+\delta\phi_{3/2}$, respectively. In the above
expressions, $\bar\Omega_{ls}^d$ is the average fraction of dark
energy before last scattering, which is negligibly small
in the cases discussed here.

\begin{figure}[t]
\centering
\leavevmode \epsfbox{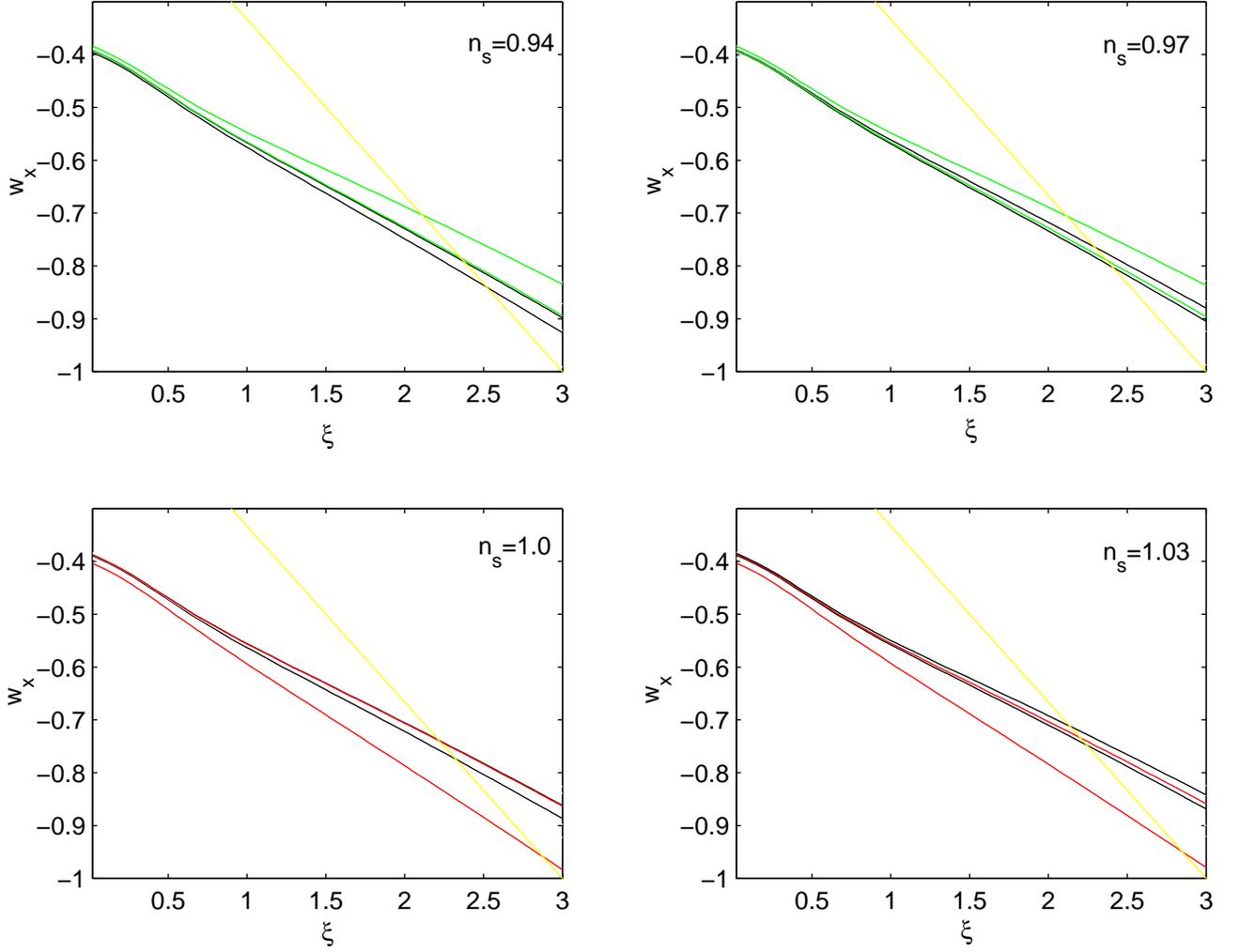}
\vskip 0.1cm
\caption{Contour Plots of the first three Doppler peaks and the first
trough location in the $(\xi, w_{x})$ plane with $\Omega_{m0}=0.2$
and $h=0.71$ for different values of $n_s$. Black, red, blue and green
lines correspond to the observational bounds on the first, second, third
peaks and the first trough, respectively (given in equations (20) and
(21)).  The upper line corresponds to the non-interacting case 
$(\xi+3w_{x}=0)$. To avoid confusion, we plot only the contours relevant 
for the admissible range. For each panel contours not shown are consistent
with this range.}
\label{figure1}
\end{figure}

\begin{figure}[t]
\centering
\leavevmode \epsfbox{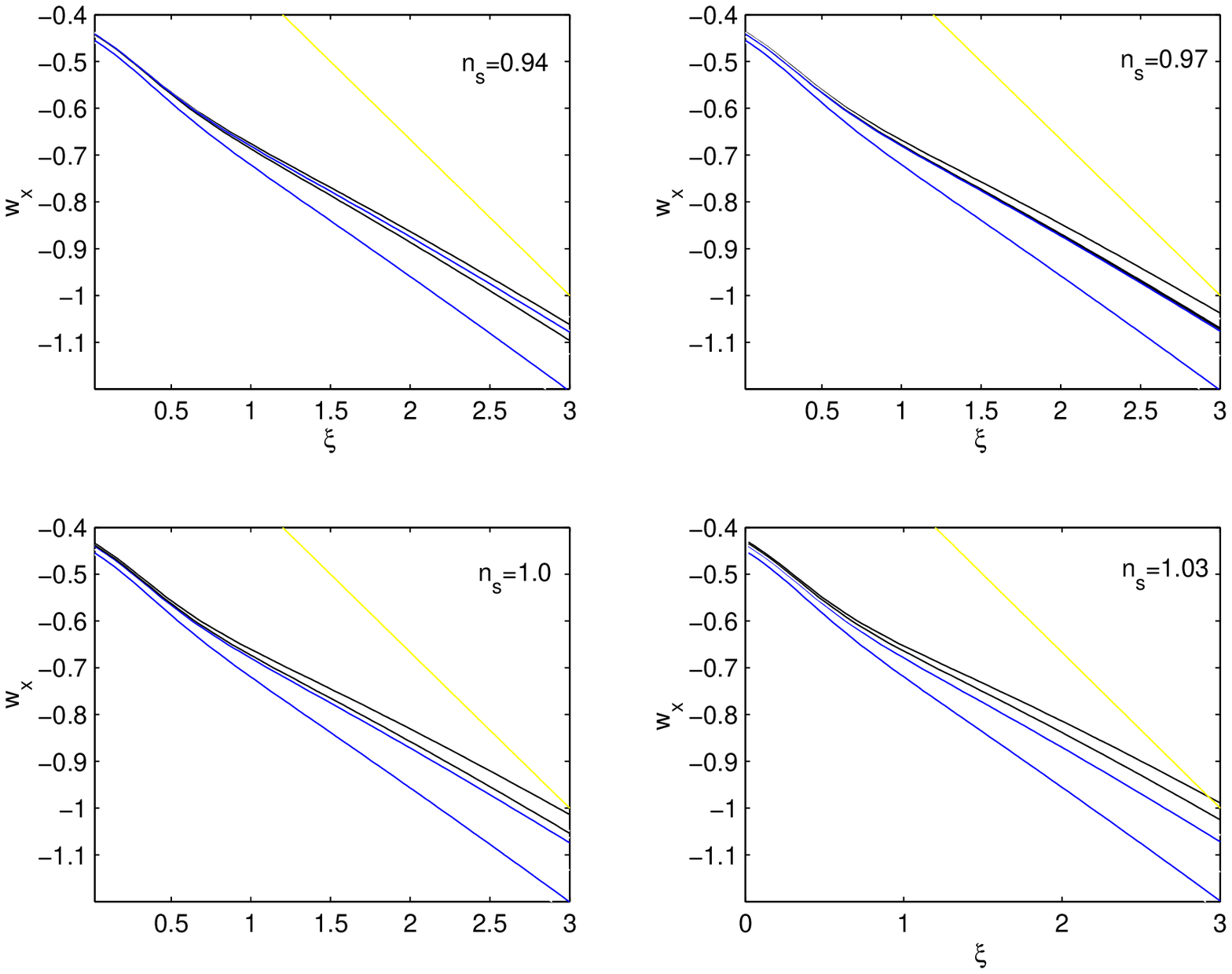}
\vskip 0.1cm
\caption{Same as Figure 1 with $h=0.71$ but $\Omega_{m0}=0.3$.}
\label{figure2}
\end{figure}

\begin{figure}[t]
\centering
\leavevmode \epsfbox{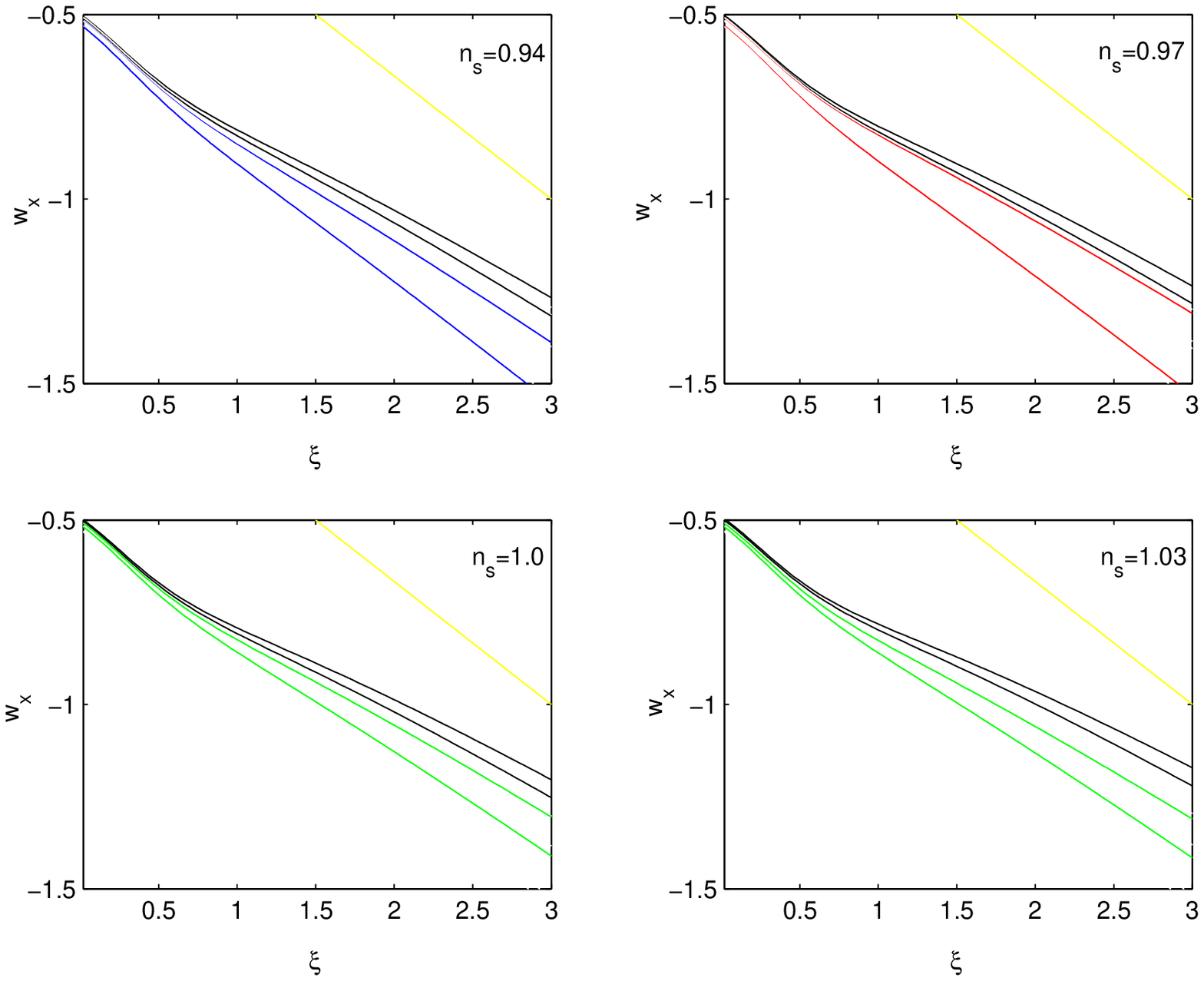}
\vskip 0.1cm
\caption{Same as Figure 1 with $h=0.71$ but $\Omega_{m0}=0.4$.}
\label{figure3}
\end{figure}

\begin{figure}[t]
\centering
\leavevmode \epsfysize=7.5cm \epsfbox{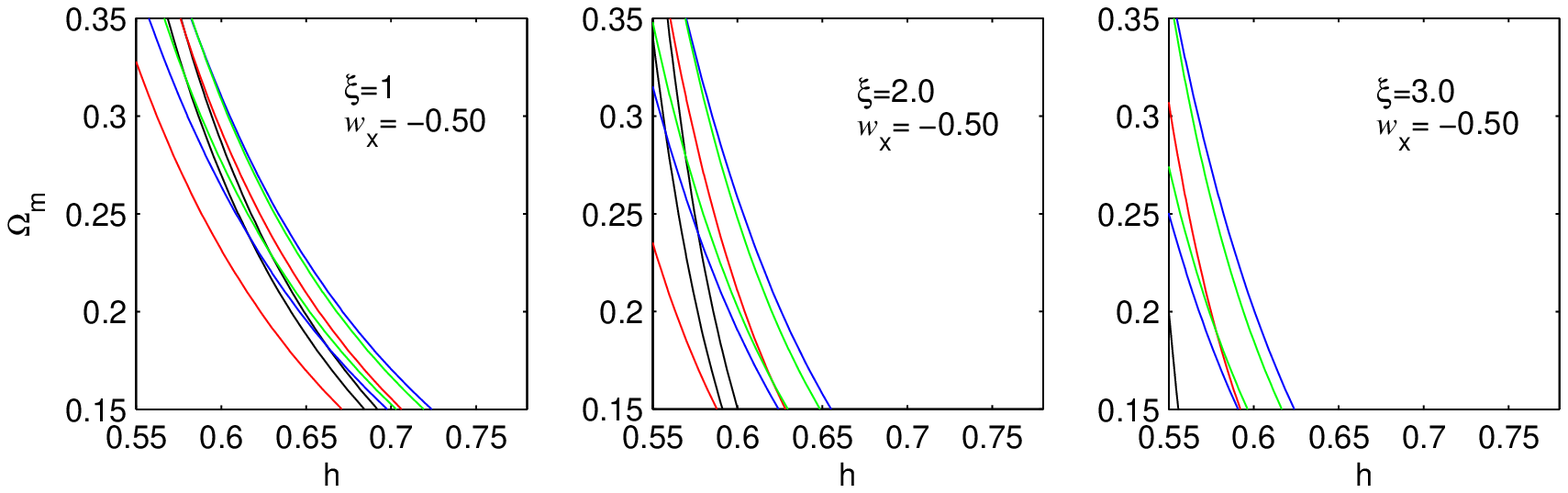}
\vskip 0.1cm
\leavevmode \epsfysize=7.5cm \epsfbox{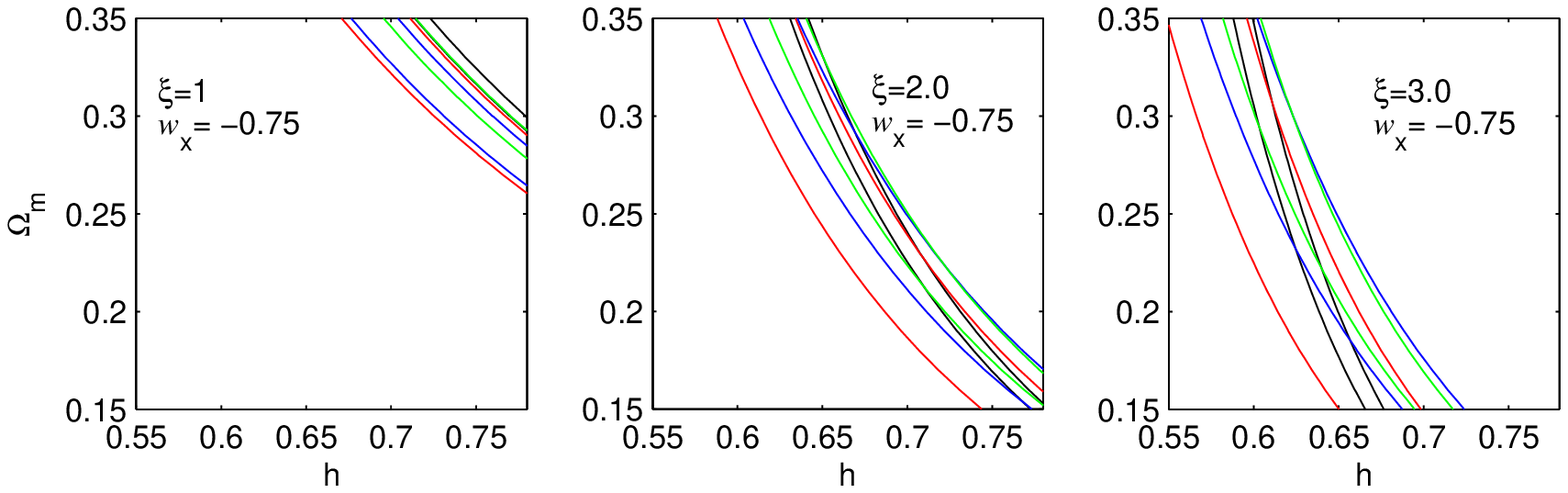}
\vskip 0.1cm
\leavevmode \epsfysize=7.5cm \epsfbox{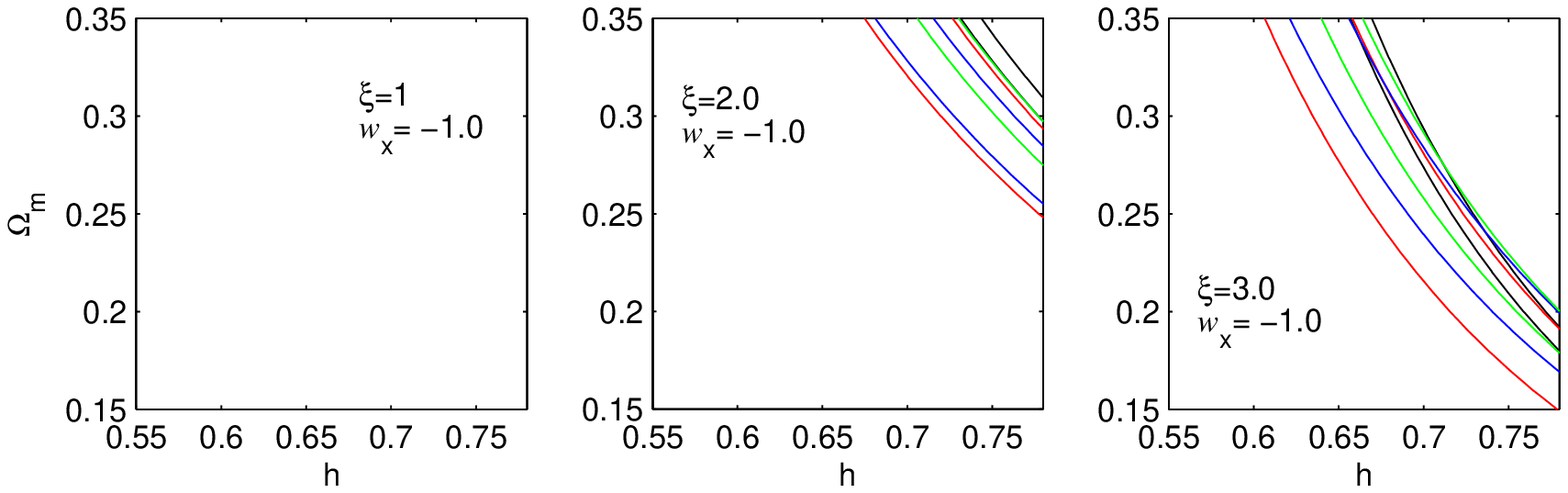}
\vskip 0.1cm
\caption{Contour Plots of the first three Doppler peaks and the first trough
location in the $(\Omega_{m0},h)$ plane with $n_s=0.97$ and for different
values of $\xi$ and  $w_{x}$. Black, red, blue and green lines correspond
to the observational bounds on the first, second, third peaks and the first 
trough,respectively (given in equations (20) and (21)).}
\label{figure4}
\end{figure}

\begin{figure}[t]
\centering
\leavevmode \epsfysize=8cm \epsfbox{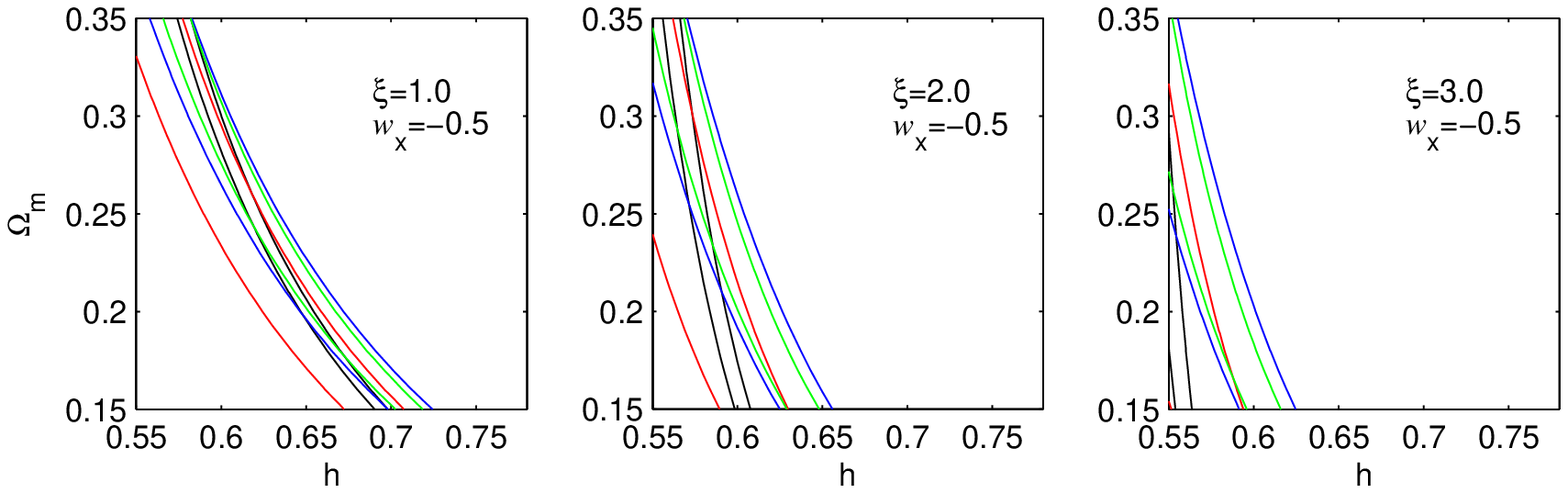}
\vskip 0.1cm
\leavevmode \epsfysize=8cm \epsfbox{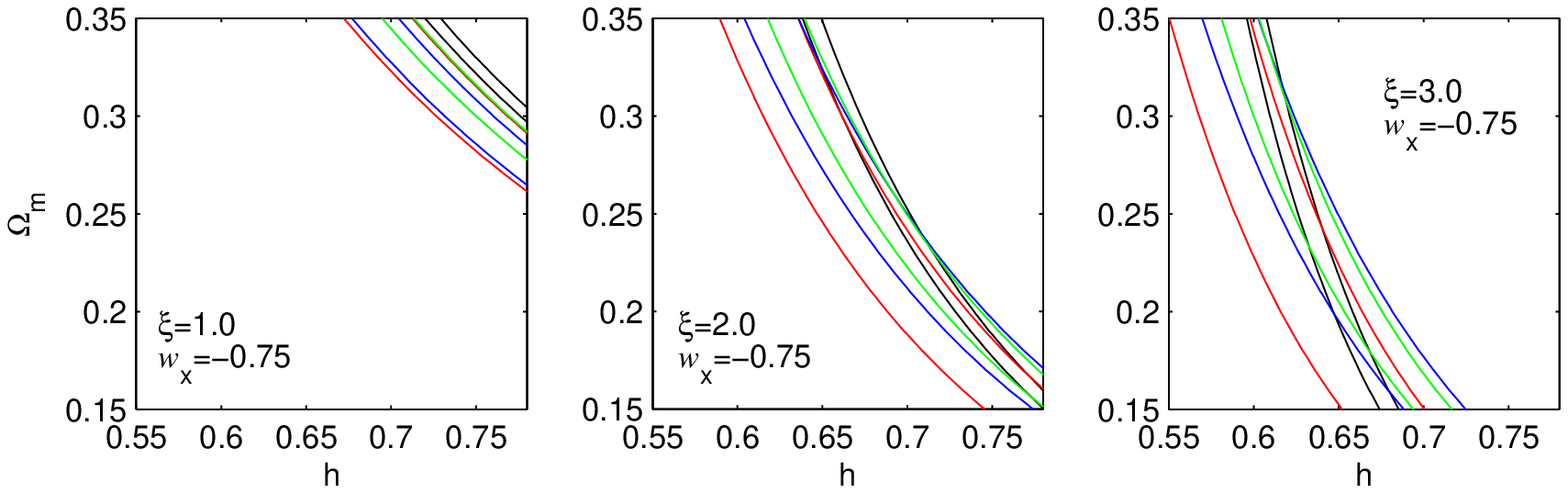}
\vskip 0.1cm
\leavevmode \epsfysize=8cm \epsfbox{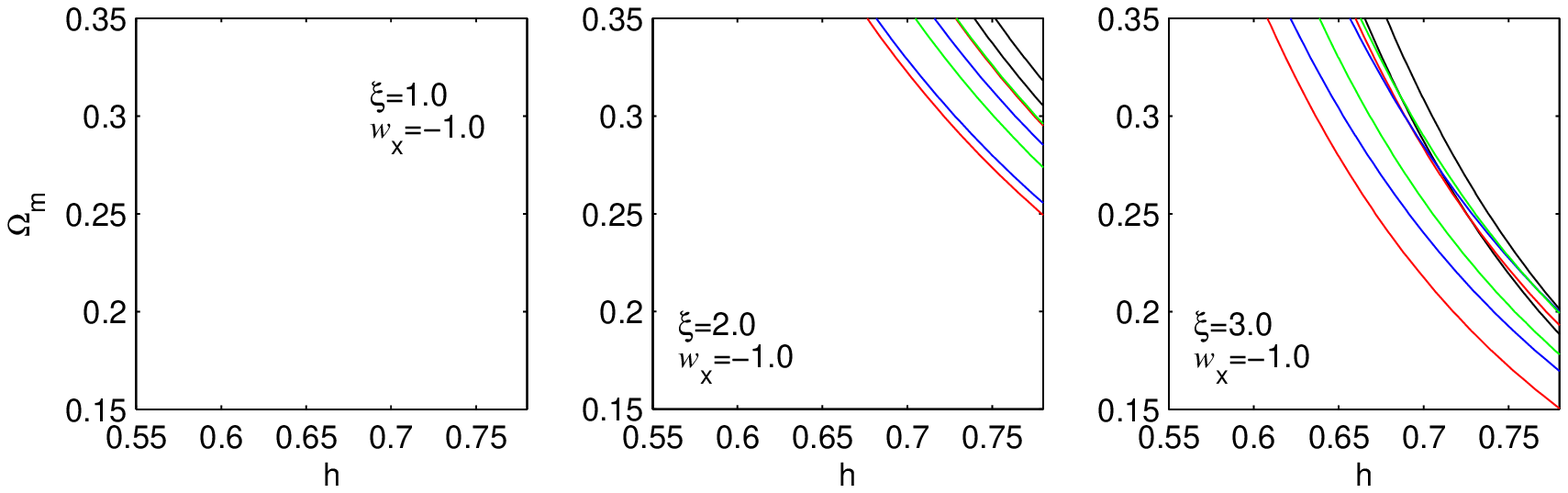}
\vskip 0.1cm
\caption{Same as Figure 4 but with $n_s=1.0$ and different values of $\xi$ and
$w_x$.}
\label{figure5}
\end{figure}


\begin{thebibliography}{99}
\bibitem{cmbsuper} Bernadis, P. De. et al., 2000, {\it Nature} {\bf
404}, 955; Hanany, S. et al., 2000, {\it Astrophys. J.} {\bf 545}, L5;
Balbi, A. et al., 2000, {\it Astrophys. J.} {\bf 545}, L1; 
{\em ibid.}, 2001 {\bf 558} L145;
Perlmutter, S. et al., 1997, {\it Astrophys. J.} {\bf 483}, 565;
Perlmutter, S.  et al., 1998, {\it Nature}, {\bf 391}, 51; Garnavich, P.M.
et al., 1998, {\it Astrophys. J.} {\bf 493}, L53; 
Riess, A.G. 1998, {\it Astron. J.} {\bf 116}, 1009.
\bibitem{diff} Caldwell, R.R., Dave, R. and Steinhardt, P.J., 1998,
{\it Phys. Rev. Lett.} {\bf 80}, 1582;
Peebles, P.J.E. and Ratra, B., 1988, {\it Astrophys. J.} {\bf325}, L17;
Steinhardt, P.J., Wang, L., and Zlatev, I., 1999, {\it Phys. Rev. Lett.}
{\bf 59}, 123504;
Uzan, J.P., 1999, {\it Phys. Rev. D} {\bf 59}, 123510; 
Amendola, L. 2000, {\it Phys. Rev. D}, {\bf 62}, 043511; 
Sen, S. and  Sen, A.A.,  2001a, {\it Phys. Rev. D} {\bf 63}, 124006; 
Sen, A.A. and Sen, S. 2001b, {\it Mod. Phys. Lett. A}, {\bf 16}, 1303;
Banerjee, N. and Pav\'{o}n, D., 2001, {\it Phys. Rev. D} {\bf 63}, 043504; 
Banerjee, N. and Pav\'{o}n, D., 2001, {\it Class. Quantum Grav.} {\bf 18}, 593.
\bibitem{stein} Steinhardt, P.J., in {\em Critical Problems in Physics},
edited by Fitch, V.L. and Marlow, D.R. (Princeton University Press,
Princeton, N.J., 1997).
\bibitem{anek} Zlatev, I., Wang, L., and Steinhardt, P.J., 1999,
{\it Phys. Rev. Lett.} {\bf 82}, 896; 
Zlatev,I. and Steinhardt, P.J., 1999, {\it Phys. Lett. B} {\bf 459}, 570;
Arkani--Hamed, N., Hall, L.J., Kolda, C.F., and Murayama, H., 2000,
{\it Phys. Rev. Lett.} {\bf 85}, 4434;
Griest, K., 2002, {{\it Phys. Rev. D}} {\bf 66}, 123501;
Cohen, I. astro-ph/0304029; 
Pietroni, M., 2003, {{\it Phys. Rev. D}} {\bf 67}, 103523.
\bibitem{diego1} Chimento, L.P., Jakubi, A.S., Pav\'{o}n, D. and Zimdahl,
W., 2003, {\it Phys. Rev. D} {\bf 67}, 083513; 
Chimento, L.P., Jakubi, A.S. and  Pav\'{o}n, D., 2003, {\it Phys.
Rev. D} {\bf 67}, 087302.
\bibitem{Dalal} Dalal, N., Abazajian, K., Jenkins, E. and Manohar, A.V., 2001,
{{\it Phys. Rev. Lett.}} {\bf 87}, 141302.
\bibitem{scaling} Zimdahl, W. and Pav\'{o}n, D., 2003, {\it Gen. Rel.
Grav.} {\bf 35}, 413.
\bibitem{plb} Zimdahl, W., Pav\'{o}n, D. and Chimento, L.P., 2001,
{{\it Phys. Lett. B}} {\bf 521}, 133.
\bibitem{perl}Perlmutter, S. et al. 1999, {\it Astrophys. J.}, {\bf 517}, 565;
\bibitem{wmap} Spergel, D.N. et al., 2003 {\it Astrophys. J. Suppl.} 
{\bf 148}, 175; 
Page, L. et al., LANL preprint astro-ph/0302220, {\it Astrophys. J.} (in the press).
\bibitem{boom} Ruhl,J.E. et al.,2003, {\it Astrophys. J.} {\bf 599}, 786.        
\bibitem{card} Sen, A.A. and Sen, S., 2003, {\it Phys.Rev.D} {\bf 68}, 023513;
Sen, S. and Sen,A.A., 2003, {\it Astrophys. J}, {\bf 588}, 1; Bento, M.C., 
Bertolami, O., and Sen,A.A., 2003, {\it Phys. Lett. B} {\bf 575}, 172.
\bibitem{hu} Hu, W., Fukugita. M., Zaldarriaga, M. and Tegmark., M., 2001,
{\it Astrophys. J.} {\bf 549}, 669.
\bibitem{doran1} Doran, M., Lilley, M.J., Schwindt, J. and Wetterich, C., 2001,
{\it Astrophys. J.} {\bf 559}, 501.
\bibitem{doran2}Doran, M. and Lilley,M.J., 2002, {\it Mon. Not. Roy. Astron. Soc.}
{\bf 330} 965.
\bibitem{tonry} Tonry, J.L. et al., 2003, {{\it Astrophys. J.}} {\bf 594}, 1
\bibitem{barris} Barris, B.J. et al., 2003 {\it astro-ph}/0310843, 
{\it Astrophys. J.} (in the press).
\bibitem{melchiorri} Melchiorri, A. and \"{O}dman, C.J., 2003, {\it
Phys. Rev. D} {\bf 67}, 081302.
\bibitem{paris} Brax, P., Martin, J. and Uzan, J.P. (editors)
Proceedings of the IAP Conference ``On the Nature of Dark Energy"
(Frontier Group, Paris, 2003).
\bibitem{param} Durrer, R., Novosyadlyl, D. and Apunevych, S., 2003,
{\it Astrophys. J.} {\bf 583}, 33.
\end{thebibliography}
\end{document}